\documentclass[prd,twocolumn,aps,amsmath,amssymb,nofootinbib,preprintnumbers]{revtex4}

\usepackage{graphicx}
\usepackage{dcolumn}
\usepackage{bm}
\usepackage{amsmath}
\usepackage{amsfonts}

\def\drawbox#1#2{\hrule height#2pt
        \hbox{\vrule width#2pt height#1pt \kern#1pt
              \vrule width#2pt}
              \hrule height#2pt}

\def\Asym#1#2{\vcenter{\vbox{\drawbox{#1}{#2}
              \kern-#2pt       
              \drawbox{#1}{#2}}}}

\newcommand{\beq}{\begin{eqnarray}}
\newcommand{\eeq}{\end{eqnarray}}

\begin{document}

\title{Gravitino production from reheating in split supersymmetry}
\author{Rouzbeh Allahverdi$^{1}$}
\author{Asko Jokinen$^{2}$}
\author{Anupam Mazumdar$^{2}$}
\affiliation{$^{1}$~Theory Group, TRIUMF, 4004 Wesbrook Mall, Vancouver, BC, 
V6T 2A3, Canada. \\ 
$^{2}$~NORDITA, Blegdamsvej-17, Copenhagen-2100, Denmark.}

\begin{abstract}
We discuss gravitino production from reheating in models where the
splitting between particle and sparticle masses can be larger than TeV,
as naturally arising in the context of split supersymmetry. We show
that such a production typically dominates over thermal contributions
arising from the interactions of gauginos, squarks and sleptons. We
constrain the supersymmetry breaking scale of the relevant sector for
a given reheat temperature. However the situation changes when the
gravitinos dominate the Universe and decay before nucleosynthesis. We
briefly describe prospects for a successful baryogenesis and a viable
neutralino dark matter in this case.
\end{abstract}
\preprint{NORDITA-2004-80}
\maketitle

\section{Introduction}

The recent satellite based experiments strongly favor primordial
inflation~\cite{WMAP}. Inflation is an attractive mechanism which
explains the homogeneity and the flatness problem, the large scale
structures through primordial density perturbations and the tiny
fluctuation in the cosmic microwave background
radiation~\cite{inf}. However inflation leaves the Universe cold and
devoid of any entropy. After inflation the Universe must be reheated
in order to keep the successes of the hot big bang
nucleosynthesis~\cite{BBN}.

Inspite of the phenomenal success it has been extremely hard to pin
down the inflaton sector~\cite{Lyth}. This is mainly due to our
ignorance of the physics beyond the Standard model (SM). The most
popular paradigm is the minimal supersymmetric model beyond the
SM~\footnote{Inflaton cannot be a gauge invariant flat direction of a
minimal supersymmetric SM, where supersymmetry breaking scale in the
observable sector is around $1$~TeV,
see~\cite{Jokinen}.}. Supersymmetry doubles the SM degrees of freedom
by introducing a boson known as sfermion for every fermion. In this
regard supersymmetry is a novel tool to probe the early Universe,
which has a potential to be tested in the collider
experiments. Supersymmetry is well motivated from a theoretical point
of view, if it were broken in the observable sector at the electroweak
scale, in which case it can address host of interesting issues, such
as the hierarchy between the Planck and the electroweak scale,
ameliorating the cosmological constant problem by $64$ orders of
magnitude, leading to a gauge unification at the grand unified scale,
and along with R-parity providing a stable particle known as the
lightest supersymmetric particle (LSP), which can be a suitable
candidate for the cold dark matter.

Recently there has been an interesting proposal for an intermediate
scale supersymmetry breaking at a scale above the electroweak but
below the Planck scale by splitting the masses of the fermions and the
bosons~\cite{split,adgr}. In this new scheme the bosons are heavier
than the fermions.  Although such a scheme does not attempt to address
the hierarchy problem, but it keeps the gauge unification as a
building block, and removes flavor and CP violating effects induced by
the light scalars at one loop level.  Running of the gauge couplings
require the gauginos to be light at scales close to $1-100$~TeV, while
spontaneous breaking of the electroweak symmetry requires the lightest
Higgs to be around ${\cal O}(100)$~GeV.

A priori there is no fundamental theory which fixes this scale, but
cosmological observations place severe constraint on the intermediate
scale of supersymmetry breaking. The theory also permits a light and
long lived gluino. The overproduction of gluinos place a severe bound
on the scale of supersymmetry breaking which has to be less than
$10^{13}$~GeV.

Embedding supersymmetry in gravity also leads to a new particle, which
is also fairly long lived and known as gravitino, a superpartner of
graviton. The main aim of this paper is to show that the production of
gravitinos is inevitable from a sector which is responsible for
reheating or generating entropy in the Universe.  Our analysis is very
general and it is applicable to many distinct cases, such as gravitino
production from the decay of inflaton, supersymmetric flat directions,
Q-balls, right handed Majorana sneutrino condensate, etc. All of them
are responsible for generating entropy at various stages of the
evolution of the Universe.


\section{Various sources for entropy generation}

Inflaton sector is the most prominent source for entropy
production. Assuming that the inflaton decay products thermalize
instantly, then the reheat temperature of the Universe is given by
$T_{\rm R}\approx\sqrt{\Gamma M_{\rm P}}$, where $\Gamma$  is the
inflaton decay rate to the light fermions. Inflaton can decay
perturbatively~\cite{nop,aem,kyy,ad4} and non-perturbatively into
gravitinos~\cite{non-pert}. Gravitinos can also be generated from a
thermal bath created by the inflaton decay products, mainly through
interactions of the ordinary sparticles~\cite{thermal}.

Besides inflaton there are other sources of entropy production,
supersymmetry has many flat directions, made up of gauge invariant
combinations of squarks and sleptons, which may acquire non-vanishing
vacuum expectation values (vevs) during inflation, thereby forming
homogeneous zero-mode condensates. The condensates may play a
significant role in many cosmological phenomena~\cite{Enqvist}, such
as generating baryons and dark matter particles by first fragmenting
into $Q$-balls which then decay \cite{qball} through surface
evaporation to generate late entropy, it has also been suggested that
the origin of all matter and density perturbations could, in
principle, be due to such flat directions \cite{denspert}.  The
supersymmetric flat direction also leads to the excitation of
primordial magnetic field~\cite{mag} and plays important role in our
understanding of reheating/preheating~\cite{averdi}.

Quite similar conclusions hold for a heavy Majorana sneutrino
condensate, whose decay can generate lepton asymmetry and
entropy~\cite{many}.  For our purposes we will study the decay rate of
a generic condensate which is responsible for reheating the Universe
and then we will discuss various consequences.

\section{Gravitino production from reheating}

We denote the mass difference between a scalar field $\phi$, whose
decay reheats the Universe, and its fermionic partner ${\tilde \phi}$
by $m\equiv m_{\phi}-m_{\tilde\phi}$. The supersymmetric conserving
mass of this multiplet is assumed to be $M$, given by the
superpotential term,
\beq
W= \frac{1}{2}M\Phi\Phi +...\,,
\eeq
where $\Phi$ is the chiral superfield whose scalar component is
$\phi$.  Here $\phi$ can be the inflaton, supersymmetric flat
direction, or whatever field whose decay generates entropy.  Note that
such a mass difference naturally arises after supersymmetry breaking
from the soft mass term and B term. In addition to the $\phi$
multiplet, we define $\tilde m$ to be the the mass difference between
the SM particles and their superpartners.


In the context of split supersymmetry, it is natural to expect that $m
\gg 1$~TeV. If $\phi$ is the inflaton, $m \leq 10^{13}$ GeV will be
required from the bound on scalar and tensor perturbations~\cite{inf}
\footnote{This is strictly correct for a
single field chaotic type inflationary models.}.  It is interesting
that this bound coincides with that of the mass difference between the
SM fermions and their scalar partners, denoted by ${\tilde m}$,
derived from the requirement that gluino lifetime is less than the age
of the universe \cite{split}. So long as $m > m_{3/2}$, with $m_{3/2}$
being the gravitino mass, the process $\phi \rightarrow {\tilde \phi}
+ {\rm gravitino}$ will be kinematically allowed. Moreover, for $m >
{\rm few} \times m_{3/2}$, helicity $\pm 1/2$ gravitinos will be
mainly produced. These states essentially interact like the Goldstino
$\psi$ and the relevant couplings are \cite{goldstino}
\beq \label{lagr}
{\cal L} \supset {m^2_{\phi} - m^2_{\tilde \phi} \over \sqrt{3} m_{3/2} 
M_{\rm P}} {\phi}^* {\bar \psi} \left({1 + \gamma_5 \over 2}\right) {\tilde 
\phi} + {\rm h.c.}\,,
\eeq
leading to the partial decay width
\beq \label{gravwidth}
\Gamma_{\rm part} \simeq {1 \over 48 \pi} 
{(m^2_{\phi} - m^2_{\tilde \phi})^4 \over m^2_{3/2} M^2_{\rm P} M^3}\,.
\eeq
Here $M_{\rm P} = 2.4 \times 10^{18}$~GeV is the reduced Planck
mass. The number of gravitinos produced per $\phi$ decay will be given
by $\Gamma_{\rm part} {\Gamma}_{\rm tot}^{-1}$, where $\Gamma_{\rm
tot} = \left(g_* \pi^2/30 \right)^{1/2} T^2_{\rm R}/M_{\rm P}$ is the
total decay rate of $\phi$ (and ${\tilde \phi}$ if $m \ll M$). Here
$T_{\rm R}$ denotes the reheat temperature of the Universe and $g_*$
is the number of relativistic degrees of freedom at $T_{\rm R}$. If
$T_{\rm R} > {\tilde m}$, we have $g_* = 225$. For $T_{\rm R} <
{\tilde m}$ squarks and sleptons are decoupled from the thermal bath
but this will be numerically irrelevant for our calculations. If $M
\gg m$, which we consider to be the case and after taking into account
of the dilution factor, $3 T_{\rm R}/4 m_{\phi}$, we find
\beq \label{phidecay}
\left({n_{3/2} \over s}\right)_{\phi} \simeq 2.5 \times 10^{-2} 
{m^4 \over m^2_{3/2} T_{\rm R} M_{\rm P}}\,.
\eeq
The most interesting point is that $M$ drops out of the
calculations. In an opposite limit $M \ll m$, which happens for squark
and slepton fields in this scenario, see Refs.~\cite{split,adgr}, the
result will be smaller by a factor of $16$.

There are two other sources of gravitino production from ordinary
sparticles in the early Universe. One is through the scatterings of
gauge and gaugino quanta in the primordial thermal bath. This is most
effective when the bath has its highest temperature, $T_{\rm R}$,
leading to \cite{thermal}
\beq \label{scattering}
\left({n_{3/2} \over s}\right)_{\rm sc} \simeq \left(1 + {M^2_{\tilde g} \over 
12 m^2_{3/2}}\right) \left({T_{\rm R} \over 10^{10}~{\rm GeV}}\right) \times 
10^{-12}\,,
\eeq
where $M_{\tilde g}$ is the gluino mass. Gravitinos are also produced
in the decay of ordinary sparticles. If all sparticles have thermal
equilibrium abundance, we have~\cite{goldstino}
\beq \label{decay}
\left({n_{3/2} \over s}\right)_{\rm dec} \simeq 
\left({{\tilde m} \over m_{3/2}}
\right)^2 \left({{\tilde m} \over 10^9 ~ {\rm GeV}}\right) \times 10^{-13}\,.
\eeq
We remind that ${\tilde m}$ is the mass difference between the
SM particles and their super partners. The contribution from
sparticle decays dominates when~\cite{adgr}
\beq \label{deccond}
{T_{\rm R}}^{1/3} {m_{3/2}}^{2/3} < {\tilde m} < T_{\rm R}\,.
\eeq
Outside this range, contribution from scatterings will be dominant.         

Depending on whether gravitino is the LSP or not, its abundance is
constrained by various considerations. If gravitino is not the LSP, it
will be unstable with a decay lifetime, $\tau_{3/2} \sim M^2_{\rm
P}/m^3_{3/2}$. For $m_{3/2} > 100$~GeV gravitino decays before BBN,
and hence does not affect the late cosmology.  Its decay, however,
will produce one neutralino per gravitino. If gravitino decay occurs
below the neutralino freeze-out temperature, non-thermal LSPs thus
produced should not overclose the Universe. This implies that,
\beq \label{nlspbound}
{n_{\chi} \over s} \leq 3 \times 10^{-10} \left({1~{\rm GeV} \over m_{\chi}} 
\right)\,,
\eeq
provided that neutralino annihilation is not efficient at the time of
gravitino decay, where we have denoted the neutralino mass by
$m_{\chi}$. If $m_{3/2} < 100$~TeV, gravitino lifetime is long enough
to affect nucleosynthesis. For $1~{\rm TeV} \leq m_{3/2} \leq 100$
TeV, hadronic decay modes lead to the strongest constraints
\cite{kkm}, while, for $m_{3/2} < 1$ TeV, radiative decays yield the
most stringent bounds~\cite{cefo}.

Finally, if gravitino is the LSP, its abundance should not exceed that
of a dark matter contribution. Thus,
\beq \label{lspbound}
{n_{3/2} \over s} \leq 3 \times 10^{-10} \left({1~{\rm GeV} \over m_{3/2}} 
\right)\,.
\eeq

The total abundance of gravitinos in our case is given by
\beq \label{total}
\left({n_{3/2} \over s}\right)_{\rm tot} = \left({n_{3/2} \over s}\right)_{
\phi} + \left({n_{3/2} \over s}\right)_{\rm dec} + \left({n_{3/2} 
\over s} \right)_{\rm sca}\,.
\eeq
We now require that the contribution from $\phi$ decay to be
subdominant, so that the constraints derived in Ref.~\cite{adgr}
remain valid. If the dominant contribution to the gravitino abundance
comes from scatterings, see Eq.~(\ref{scattering}), and by assuming
that $M_{\tilde g} \sim m_{3/2}$, this will require,
\beq \label{scadom}
m^4 < 10^{-2} m^2_{3/2} T^2_{\rm R}\,.
\eeq
This should particularly hold when $T_{\rm R} < {\tilde m}$, which
leads to a tighter bound, $m^2 < 0.1 m_{3/2} {\tilde m}$. In the
opposite case, when sparticle decays, contribution from
Eq.~(\ref{decay}) dominates the gravitino abundance, we find the bound
to be,
\beq \label{decdom}
m^4 < 10^{-2} {\tilde m}^3 T_{\rm R}\,.
\eeq
An absolute upper bound, $m^2 < 0.1 {\tilde m}^3/m^2_{3/2}$, can be
obtained in this case after using the first inequality in
Eq.~(\ref{deccond}).

One comment is in order at this point. Gravitinos are fermions, and
hence their occupation number is limited by the Pauli blocking which
has to be $\leq 1$. The available phase space for $\phi$ decay
constrains the physical momentum of the produced gravitinos to be
$k_{3/2} < m$. This, as noted in \cite{ad4}, implies an upper limit
$\simeq 3 \times 10^{-4} \left(m/T_{\rm R}\right)^3$ on the comoving
abundance of gravitinos from $\phi$ decay~\footnote{Here we have
assumed the maximum occupation number throughout the available phase
space. This is a valid approximation despite the fact that $k_{3/2}$
is narrowly peaked around $m$ at the time of production. Note that
$\phi$ decay does not occur instantly, and hence during its lifetime
$k_{3/2}$ will sweep the phase space due to the Hubble expansion.}. The
quantity $n_{3/2}/s$ reaches the saturation limit for
\beq \label{satur}
m_{\rm sat} \simeq 10^{-2} {m^2_{3/2} M_{\rm P} \over T^2_{\rm R}}\,.
\eeq
When $m > m_{\rm sat}$, the left-hand side of Eq.~(\ref{phidecay})
should be replaced by $3 \times 10^{-4} \left(m/T_{\rm
R}\right)^3$. The bounds in Eqs.~(\ref{scadom}) and (\ref{decdom})
will in this case be modified accordingly.

So far we have assumed that $\phi$ dominates the energy density of the
Universe at the time of decay. Now let us consider the case in which
$\phi$ carries a fraction $r < 1$ of the total energy density when it
decays. The $\phi$ field cannot be the inflaton in this case, and
hence another field should be responsible for reheating the
Universe. The candidates are supersymmetric flat directions, sneutrino
condensate and perhaps the $Q$-balls.  If we denote the temperature of
a thermal bath at the time of $\phi$ decay by $T_{\rm d}$, the
bounds in Eqs.~(\ref{scadom}) and (\ref{decdom}) will be replaced by
\beq \label{scadom1}
m^4 < 10^{-2} r^{-1} m^2_{3/2} T_{\rm R} T_{\rm d}\,,
\eeq
and
\beq \label{decdom1}
m^4 < 10^{-2} r^{-1} {\tilde m}^3 T_{\rm d}\,,
\eeq
respectively. It is interesting to note that the constraint on $m$
does not change considerably. Even if $r = 10^{-4}$, the upper bound
on $m$ will be at most weakened by one order of magnitude. Also note
that $T_{\rm d} \ll T_{\rm R}$ can compensate for $r \ll 1$. Our
bounds in Eqs.~(\ref{scadom}) and (\ref{decdom}) are therefore
practically valid for any species which undergoes an
out-of-equilibrium decay.

In addition to the entropy production during reheating, a stage of
out-of-equilibrium decay is usually needed in supersymmetric theories
for a successful cosmological scenario. If the reheat temperature
after inflation is too high, gravitino production from the
interactions of ordinary sparticles in a thermal bath exceeds the
bounds set by nucleosynthesis (for unstable gravitino) and dark matter
(for stable gravitino). A late stage of entropy release will be
necessary in this case. In addition, supersymmetric flat directions
typically generate a large amount of baryon asymmetry via Affleck-Dine
mechanism~\cite{flat}. The dilution of this excessive asymmetry
requires late entropy release and late evaporation of $Q$-balls can
ameliorate this situation. However $Q$-ball evaporation is also a
source for gravitino production.  As mentioned earlier, any
out-of-equilibrium decay directly produces gravitinos and will
therefore be subject to the bounds coming from Eqs.~(\ref{scadom}) and
(\ref{decdom}). For a late stage of entropy release, $T_{\rm R} \ll
{\tilde m}$, a tighter bound on $m$ is expected.

\section{Cosmological consequences of dominating gravitinos}

Every stage of entropy release generates gravitinos, one alternative
paradigm could be that the gravitinos produced in $\phi$ decay
dominate the energy density of the Universe~\cite{adgr}. Note that the
situation is now slightly different from the case when gravitinos from
scatterings or decay of ordinary sparticles dominate. There gravitino
energy is essentially the same as the temperature of a thermal bath at
the time of its production.  Gravitinos then become non-relativistic
when $T \simeq m_{3/2}$. Here, however, gravitino energy is $\simeq
m$, which can be very different from $T_{\rm R}$. If $m < T_{\rm R}$,
gravitinos become non-relativistic at $T > m_{3/2}$, and hence can
dominate at an earlier time. The opposite situation will happen for $m
> T_{\rm R}$. In this case gravitino decay is responsible for the last
stage of reheating which will dilute the existing (thermal) relic
neutralinos and baryon asymmetry. This can be considered as a problem
turned into a virtue, in particular if baryon asymmetry was
(over)produced via Affleck-Dine mechanism.  However, gravitino decay
will also produce neutralinos, $\chi$, with an abundance,
\beq \label{chiabun}
{n_{\chi} \over s} \simeq \left({m_{3/2} \over M_{\rm P}}\right)^{1/2}\,.
\eeq
For $m_{3/2} > 10^5$ GeV, such that gravitino decay does not affect
nucleosynthesis, this abundance is much larger than the dark matter
bound. A large annihilation cross-section $\langle \sigma_{\chi}
v_{\rm rel} \rangle$ will therefore be needed in order to bring the
neutralino abundance down to an acceptable level. Note that $\langle
\sigma_{\chi} v_{\rm rel} \rangle = c/m^2_{\chi}$, where in the case
of split supersymmetry, $c = 3 \times 10^{-3}$ for a mostly Higgsino
$\chi$, and $c = 10^{-2}$ for a mostly Wino type neutralino,
$\chi$,~\cite{adgr}. The final abundance for $\chi$ will then be given
by,
\beq \label{finalchiabun}
{n_{\chi} \over s} \simeq {m^2_{\chi} \over c (m^3_{3/2} 
M_{\rm P})^{1/2}}\,,
\eeq
where $s \propto T^3_{3/2} \propto (m^3_{3/2}/M_{\rm P})^3$ after
gravitino decay. An interesting point is that $\chi$ abundance in this
case only depends on $m_{\chi}$ and $m_{3/2}$, and viable neutralino
dark matter determines the acceptable part of this two-dimensional
parameter space.

One can also think of a following intriguing possibility. A large
baryon asymmetry is generated via Affleck-Dine mechanism (perhaps
through $Q$-ball formation) in split supersymmetry. The Q-ball decay,
which has a longer lifetime than a homogeneous condensate, then
produces a large number of gravitinos along with other
fermions. Gravitinos eventually dominate the energy density of the
Universe and their decay sufficiently dilutes the baryon asymmetry, as
well as producing non-thermal dark matter. More detailed study of
these issues will be presented in a future publication~\cite{future}.
                  
Note that gluinos can also be produced in a similar way from the decay
of the $\phi$ field. If kinematically allowed the gravitinos can decay
into gluon and gluinos. As discussed earlier in Ref.~\cite{split}, the
gluinos can be long lived, nevertheless if their mass is above $1$~TeV
and $\tilde m\sim 10^{9}$~GeV then they decay before nucleosynthesis,
see also Ref.~\cite{nunez}. We do not consider gluino cosmology
further in this paper.

\section{Conclusion}

Under general circumstances every entropy production process is
accompanied by gravitino production. We stress that particularly in
the context of split supersymmetry such a contribution can easily
dominate over thermal generation and from the decay of ordinary
sparticles.

Our main results, given by Eqs.~(\ref{scadom},\ref{decdom}),
constrain the supersymmetry breaking scale of the sector which is
responsible for reheating the Universe. Note that the bounds are
robust because the supersymmetry conserving mass of the decaying field
does not appear in the constraints.

In order to evade these bounds one could alternatively imagine that
the unstable gravitinos were abundantly produced and dominated the
Universe. For a sufficiently massive gravitinos the late stage of
reheating does not affect nucleosynthesis. However the entropy release
would dilute baryon asymmetry created earlier and produce neutralinos.
A large annihilation of neutralinos can bring the abundance to match
the current observed value for the cold dark matter, the abundance
will depend only on neutralino and gravitino mass. Baryogenesis
scenarios based on supersymmetric flat directions in general produce
large baryon asymmetry of order one, it is then possible to dilute
their abundance required for a successful big bang nucleosynthesis.

\vskip40pt

We acknowledge discussions with Holger Nielsen. R.A thanks NORDITA and
NBI for their kind hospitality during the course of this work.


\end{document}